\documentclass[11pt]{article}
\usepackage{graphicx}
\usepackage[margin=1.25in]{geometry}
\usepackage[usenames,dvipsnames]{color}
\usepackage{url}
\usepackage[colorlinks = true,
            linkcolor = blue,
            urlcolor  = blue,
            citecolor = blue,
            anchorcolor = blue]{hyperref}

\usepackage{float}
\usepackage{physics,slashed}
\usepackage{amsfonts,amsmath,amssymb}
\usepackage{mathrsfs}
\usepackage{bm}

\usepackage{epsfig}
\usepackage{graphicx}   
\usepackage{subfigure}  
\usepackage{verbatim}   
\usepackage{color}      
\usepackage{hyperref}   
\usepackage{url}
\definecolor{nicered}{rgb}{0.5,0.,0.}
\definecolor{nicegreen}{rgb}{0.,0.5,0.}
\definecolor{niceblue}{rgb}{0.,0.,0.5}
\hypersetup{colorlinks,citecolor=nicegreen,linkcolor=nicered,urlcolor=niceblue}
\usepackage{array}
\usepackage{dcolumn}
\usepackage{multirow}
\usepackage{cprotect}
\usepackage{framed}
\usepackage{setspace}

\newcommand\pubnumber{ }
\newcommand\pubdate{ }

\newcommand{\GeV}{\textrm{GeV}}

\newcommand{\bea}{\begin{equation}\begin{aligned}}
\newcommand{\eea}{\end{aligned}\end{equation}}

\def\Title#1{\begin{center} {\LARGE #1 } \end{center}}
\def\Author#1{\begin{center}{ \sc #1} \end{center}}
\def\Address#1{\begin{center}{ \it #1} \end{center}}

\newcommand\pubblock{\rightline{\begin{tabular}{l}
\pubnumber\\
\pubdate
\end{tabular}}}
\newenvironment{Abstract}{\begin{quotation} \begin{center}
                       ABSTRACT
     \end{center}\bigskip  }{\end{quotation}}

\def\beq{\begin{equation}}
\def\eeq#1{\label{#1}\end{equation}}
\def\eeqn{\end{equation}}

\newenvironment{Eqnarray}%
   {\arraycolsep 0.14em\begin{eqnarray}}{\end{eqnarray}}
\def\beqa{\begin{Eqnarray}}
\def\eeqa#1{\label{#1}\end{Eqnarray}}
\def\eeqan{\end{Eqnarray}}

\let\bar=\overbar

\def\lsim{\mathrel{\raise.3ex\hbox{$<$\kern-.75em\lower1ex\hbox{$\sim$}}}}
\def\gsim{\mathrel{\raise.3ex\hbox{$>$\kern-.75em\lower1ex\hbox{$\sim$}}}}

\def\del{\partial}
\def\Dslash{\not{\hbox{\kern-4pt $D$}}}
\def\dslash{\not{\hbox{\kern-2pt $\del$}}}
\def\pslash{\not{\hbox{\kern-2pt $p$}}}
\def\ETmiss{\not{\hbox{\kern-4pt $E$}}_T}

\def\Dlr{\mathrel{\raise1.5ex\hbox{$\leftrightarrow$\kern-1em\lower1.5ex\hbox{$D$}}}}

\def\MSB{{\bar{M \kern -2pt S}}}
\def\msb{{\bar{\scriptsize M \kern -1pt S}}}

\def\drb{{\bar{\scriptsize D \kern -1pt R}}}

\def\GeV{{\rm GeV}}

\newcommand\snowmass{\begin{center}\rule[-0.2in]{\hsize}{0.01in}\\\rule{\hsize}{0.01in}\\
\vskip 0.1in Submitted to the  Proceedings of the US Community Study\\
on the Future of Particle Physics (Snowmass 2021)\\
\rule{\hsize}{0.01in}\\\rule[+0.2in]{\hsize}{0.01in} \end{center}}

\begin{document}

\pubblock

\Title{Probing heavy-flavor parton distribution functions at hadron colliders.}

\bigskip

\Author{Keping Xie}
\Address{ Department of Physics and Astronomy, University of Pittsburgh, Pittsburgh, PA 15260, U.S.A.}


\Author{Marco Guzzi}
\Address{Department of Physics, Kennesaw State University, Kennesaw, GA 30144, U.S.A.}


\Author{Pavel Nadolsky}
\Address{Department of Physics, Southern Methodist University, Dallas, TX 75275-0181, U.S.A.}

\begin{Abstract}
\noindent Precision measurements of heavy-flavor hadroproduction at the Large Hadron Collider (LHC) have the ability to probe heavy-flavor parton distribution functions (PDFs) in the proton.
Sensitivity of inclusive $B^\pm$ meson production cross section measurements
at LHCb 13 TeV to the bottom-quark PDF is illustrated by using theory predictions obtained with a the S-ACOT-MPS general mass variable flavor number (GMVFN) scheme within the QCD factorization formula.
This approach can easily be extended to other heavy-flavor production processes, such as $Z$ boson production in association with a bottom- or a charm-jet.
The inclusion of such measurements will represent an important improvement for future global QCD analyses that aim at reducing uncertainties of heavy-flavor PDFs, provided that a consistent general mass treatment in $pp$ collisions is used.

This work is related to the activities of the Snowmass topic groups EF03, and EF06.
\end{Abstract}

\snowmass

\def\thefootnote{\fnsymbol{footnote}}
\setcounter{footnote}{0}
\section{Introduction and Motivations}

Collinear factorization in QCD has been very successful in describing the phenomenology of a large variety of hadronic reactions, including those involving final-state hadrons whose dynamics is complicated by the presence of additional multiple scales and non-perturbative effects due to heavy-quark  (HQ) fragmentation. Assuming factorization for hadroproduction reactions, the hadronic cross section is obtained as a convolution of the partonic cross section with universal parton distribution functions (PDFs) in the proton, and fragmentation functions (FFs).
Partonic cross section contributions are calculated perturbatively and account for hard interactions at short distance, while PDFs and FFs which account for long-distance dynamics, must be determined by global analyses of experimental data~\cite{Hou:2019efy,NNPDF:2017mvq,Ball:2021leu,Bailey:2020ooq,Alekhin:2017kpj,Borsa:2022vvp,Anderle:2015lqa,Kovarik:2019xvh,Ethier:2020way}, or through lattice QCD calculations~\cite{Constantinou:2022yye,Lin:2021ukf}.
Modern global QCD analyses~\cite{Hou:2019efy,NNPDF:2017mvq,Ball:2021leu,Bailey:2020ooq,Alekhin:2017kpj} extract collinear PDFs and their combinations using deep-inelastic scattering (DIS) and fixed-target cross section measurements together with a large variety of high-precision data from the LHC, e.g., single-inclusive jet production, production of Drell-Yan pairs, top-quark pairs, and high-$p_T$ $Z$ bosons. Despite all efforts, PDFs and FFs still represent one of the major sources of uncertainty in precision calculations of standard candle observables at the LHC. In addition, heavy-flavor (HF) PDFs deserve particular attention as they are currently poorly constrained as compared to the other PDFs. Constraining HF PDFs is a twofold task. First, it corresponds to the ability of a specific QCD framework (i.e., a general mass variable flavor number (GMVFN) scheme), to give correct theory predictions for observables involving heavy quarks (HQ) when the number of quark flavors changes with energy. Second, it corresponds to the possibility of directly accessing HQ PDFs parametrized at the initial scale. This motivates our work and the use of a suitable general-mass (GM) factorization scheme, named S-ACOT-MPS~\cite{Xie:2019eoe,Xie:2021ycd}, to probe and constrain HF PDFs using high-precision data from the LHC in future global QCD analyses.
The main feature of a GMVFN scheme is that it interpolates between massless (or zero mass (ZM)) and massive (fixed-flavor number (FFN)) schemes assuming that the number of quark flavors varies with energy, and at the same time including dependence on HQ masses in relevant kinematical regions. This interpolation is achieved by introducing a subtraction mechanism to avoid double-counting in the collinear region.
The S-ACOT-MPS scheme is based on an amended version of the S-ACOT scheme developed for DIS~\cite{Aivazis:1993kh,Aivazis:1993pi,Tung:2001mv,Nadolsky:2009ge,Guzzi:2011ew}, and is applied to proton-proton collisions. It differs with respect to other available GMVFN schemes~\cite{Kniehl:2004fy,Kniehl:2005mk,Helenius:2018uul} in the treatment of the phase space.
More details about the S-ACOT-MPS scheme can be found in~\cite{Xie:2019eoe,Xie:2021ycd} and will also be presented in a forthcoming study~\cite{S-ACOT-MPS}.

In this work, we apply S-ACOT-MPS to the case of $b$-quark hadroproduction at next-to-leading order (NLO) in QCD in $pp$ collisions. We calculate theory predictions
for the $b$-quark hadroproduction cross section by including $b$-quark fragmentation contributions and compare predictions for the $p_T$ and rapidity $y$ spectra at particle level to precision measurements of $B^\pm$ meson production at LHCb at a center of mass energy of 13 TeV.

\section{Kinamtic considerations for PDFs and FFs}

\begin{figure}
\centering
\includegraphics[width=0.49\textwidth]{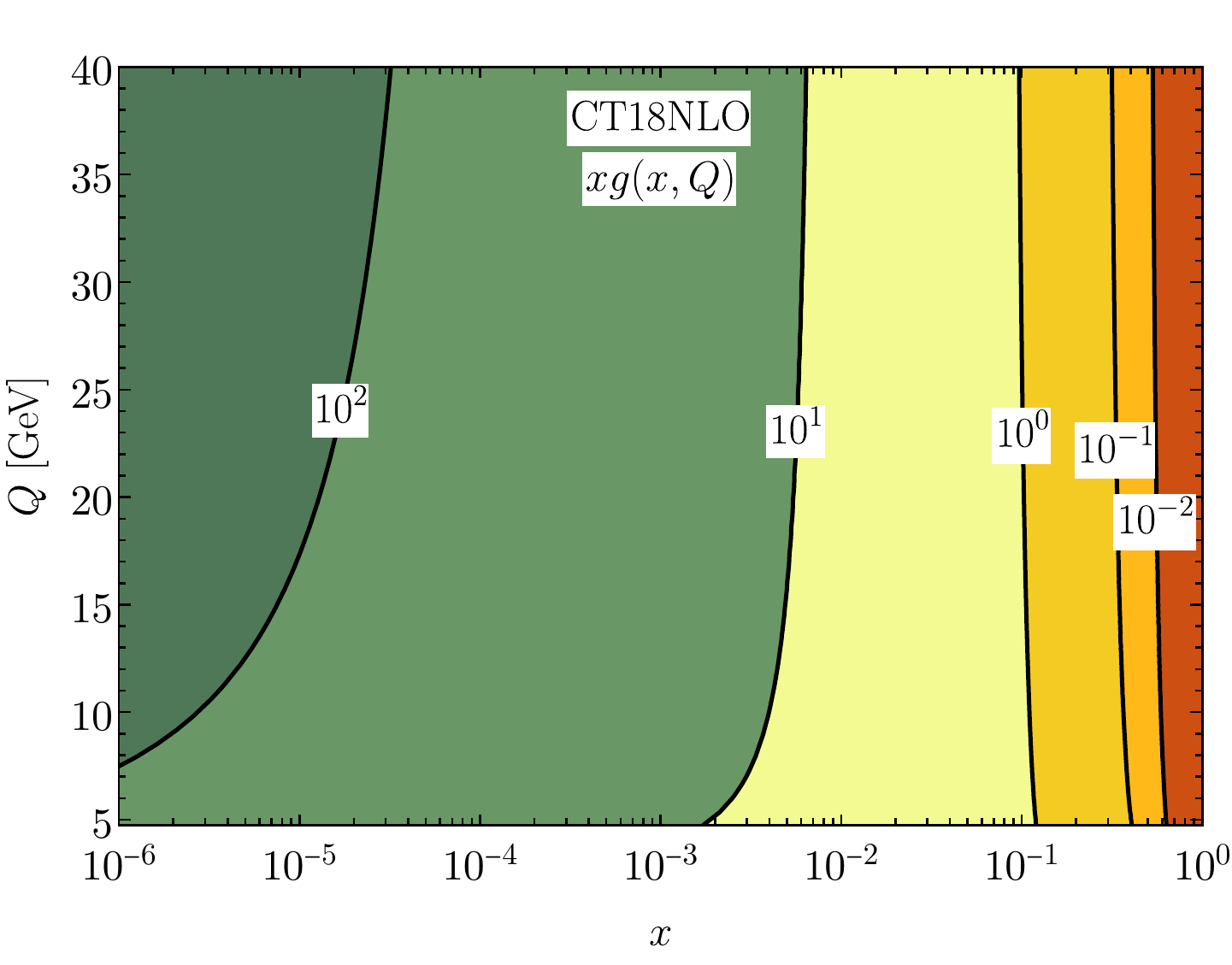}
\includegraphics[width=0.49\textwidth]{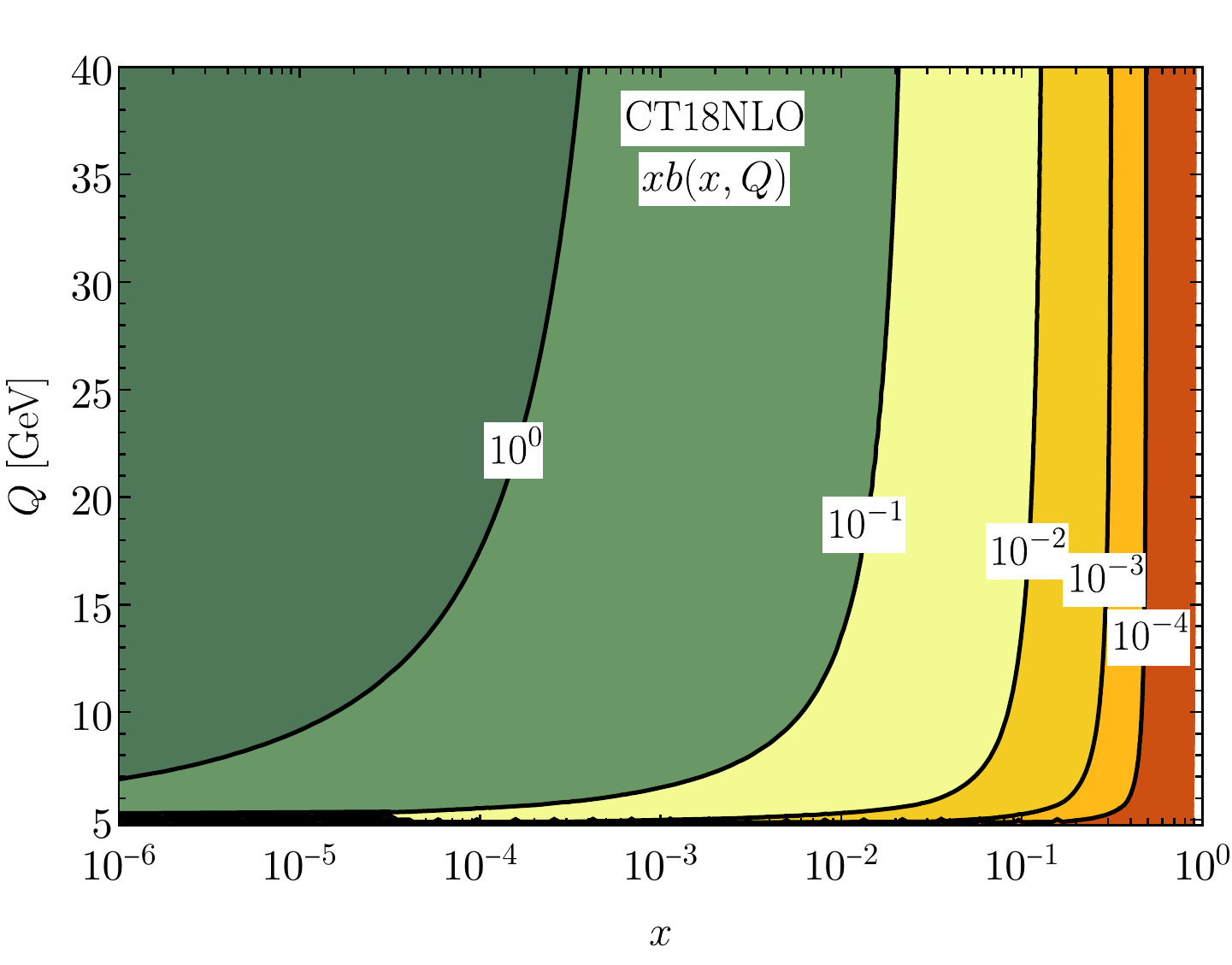}
\includegraphics[width=0.49\textwidth]{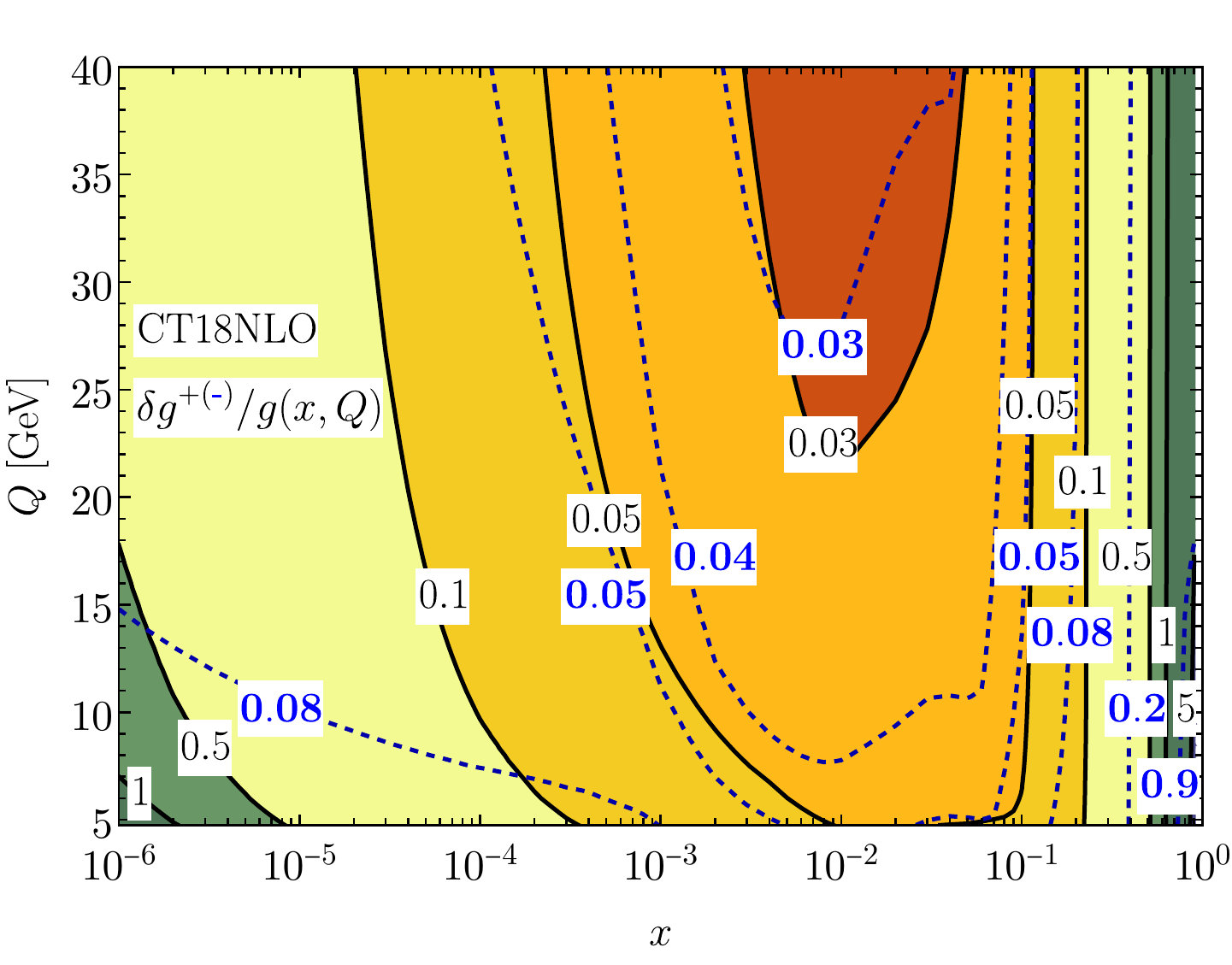}
\includegraphics[width=0.49\textwidth]{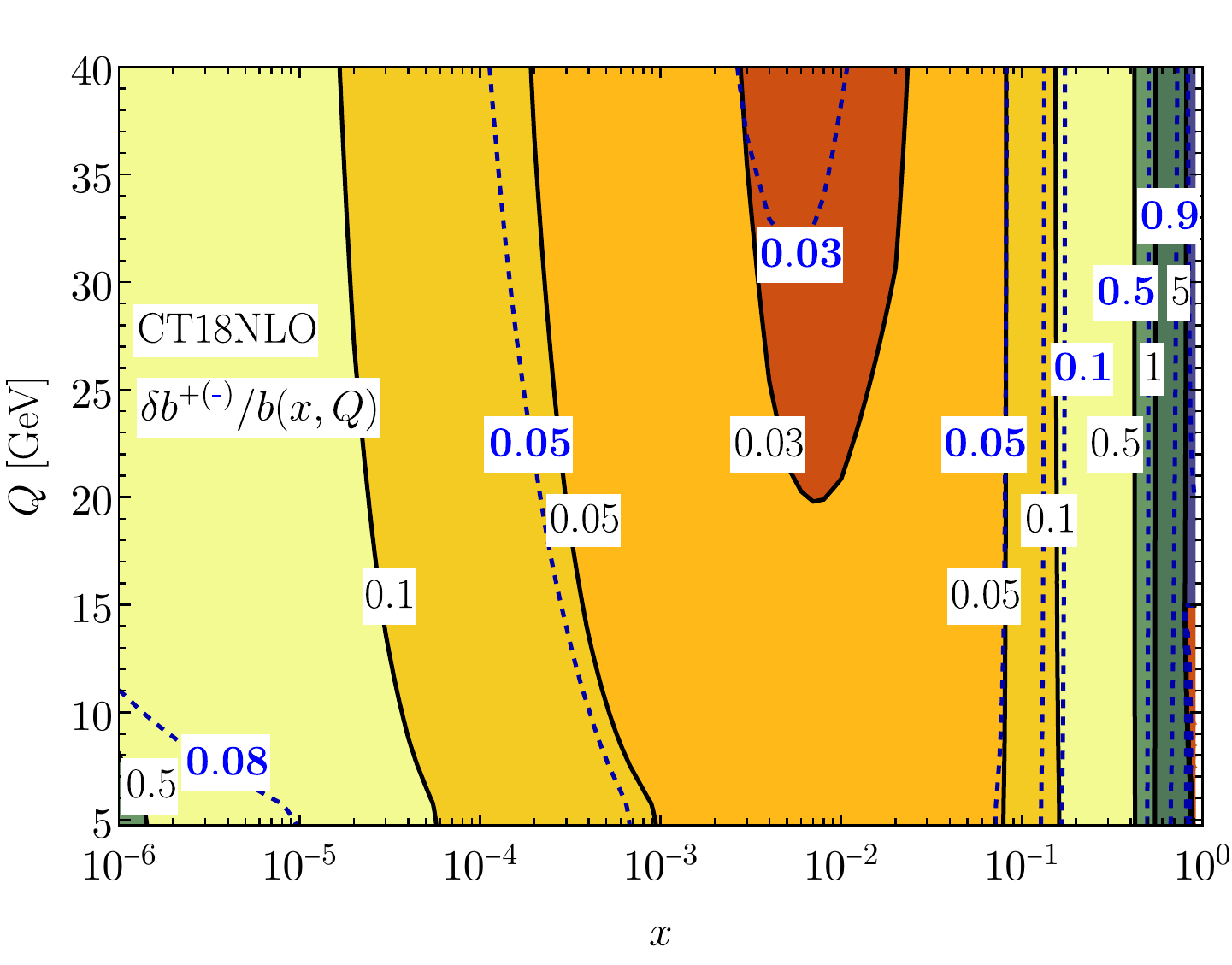}
\caption{Upper inset: Contours for the CT18NLO~\cite{Hou:2019efy} gluon(left) and $b$-quark(right) PDFs in the kinematic region sensitive to $pp\to bX$ at LHCb 13 TeV~\cite{LHCb:2017vec}. Lower insets: Corresponding PDF uncertainties obtained with the asymmetric Hessian approach~\cite{Hou:2016sho} at the 90\% Confidence Level (C.L.), with positive (negative) direction denoted as black solid (blue dashed) lines.}
\label{fig:PDF}
\end{figure}
In Fig.~\ref{fig:PDF} we illustrate the kinematic behavior of the CT18NLO~\cite{Hou:2019efy} gluon and $b$-quark densities in the $Q-x$ plane relative to inclusive $b$ (or $B$ meson) production at LHCb~\cite{LHCb:2017vec} at $\sqrt{S}=13$ TeV. LHCb detectors are able to measure properties of particles at forward rapidity $2<y<4.5$ and with transverse momentum $p_T<40~\GeV$. In this kinematic regime, PDF momentum fractions are probed down to $x\sim (Q/\sqrt{s}) e^{-y}\gtrsim 4 \cdot 10^{-6}$, where the typical scale $Q$ for the process is $Q\sim \sqrt{m_b^2+p_T^2}$.
In the upper insets of Fig.~\ref{fig:PDF} we observe a rapid growth for $xg(x,Q^2)$ and $xb(x,Q^2)$ for smaller values of $x$ and larger values of $Q$ that are represented by dark green regions. In addition, the resulting color pattern reflects strong correlation between these two PDFs becasue the $b$-quark density is radiatively generated by gluon splitting $g\to b\bar{b}$.
In the lower insets of Fig.~\ref{fig:PDF} we show the gluon and $b$-quark PDF uncertainties normalized to the respective CT18NLO best-fit. We observe that in the intermediate $x$ region at $x\sim 10^{-2}$, the PDF fractional uncertainties are approximately $3\%$ and grow rapidly as $x$ moves away from this point. In the kinematic region $x<(Q/\sqrt{s})e^{y}<0.3$ that is probed by the LHCb data, PDF uncertainties are $\mathcal{O}(10\%)$ or more.
\begin{figure}
\centering
\includegraphics[width=0.45\textwidth]{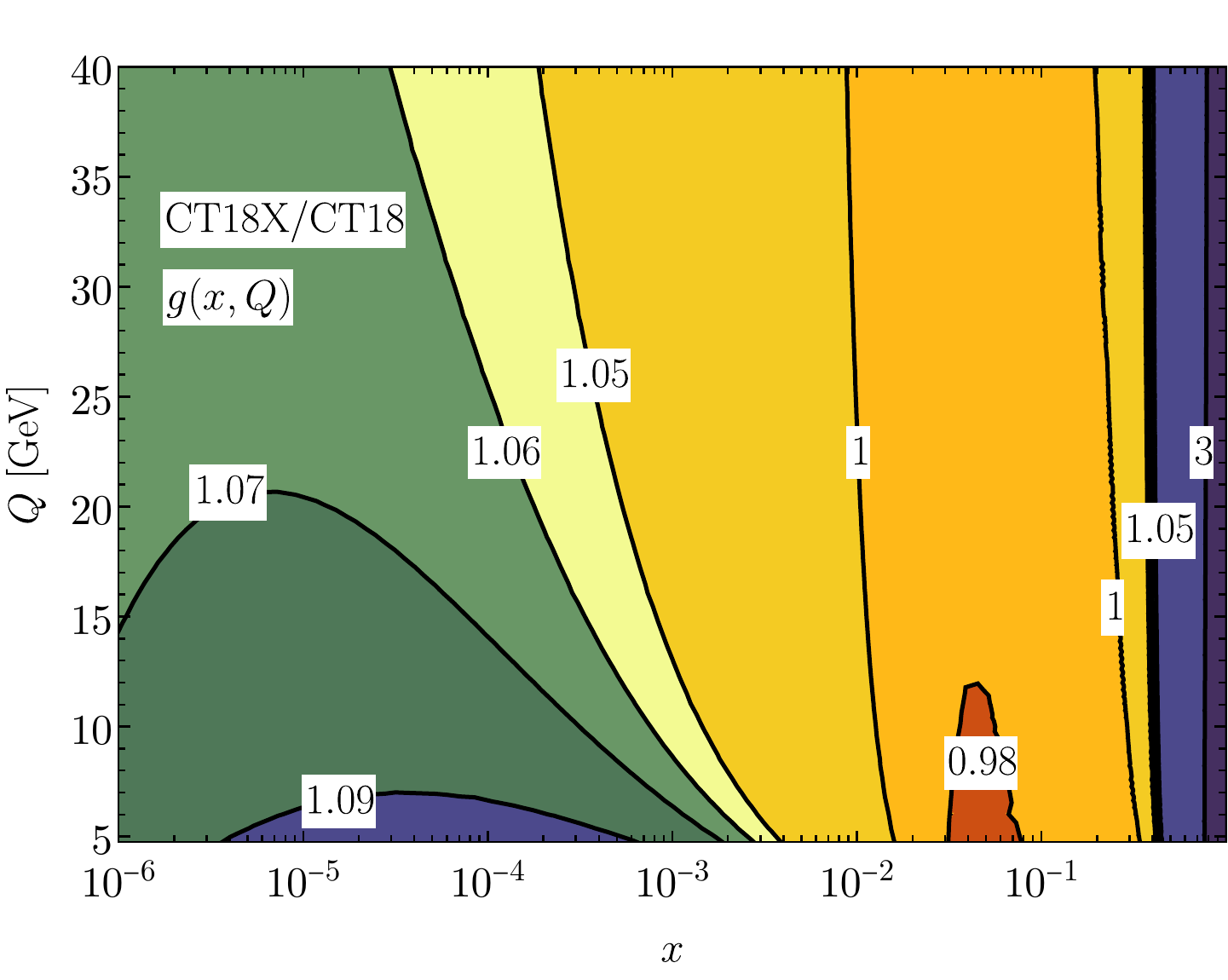}
\includegraphics[width=0.45\textwidth]{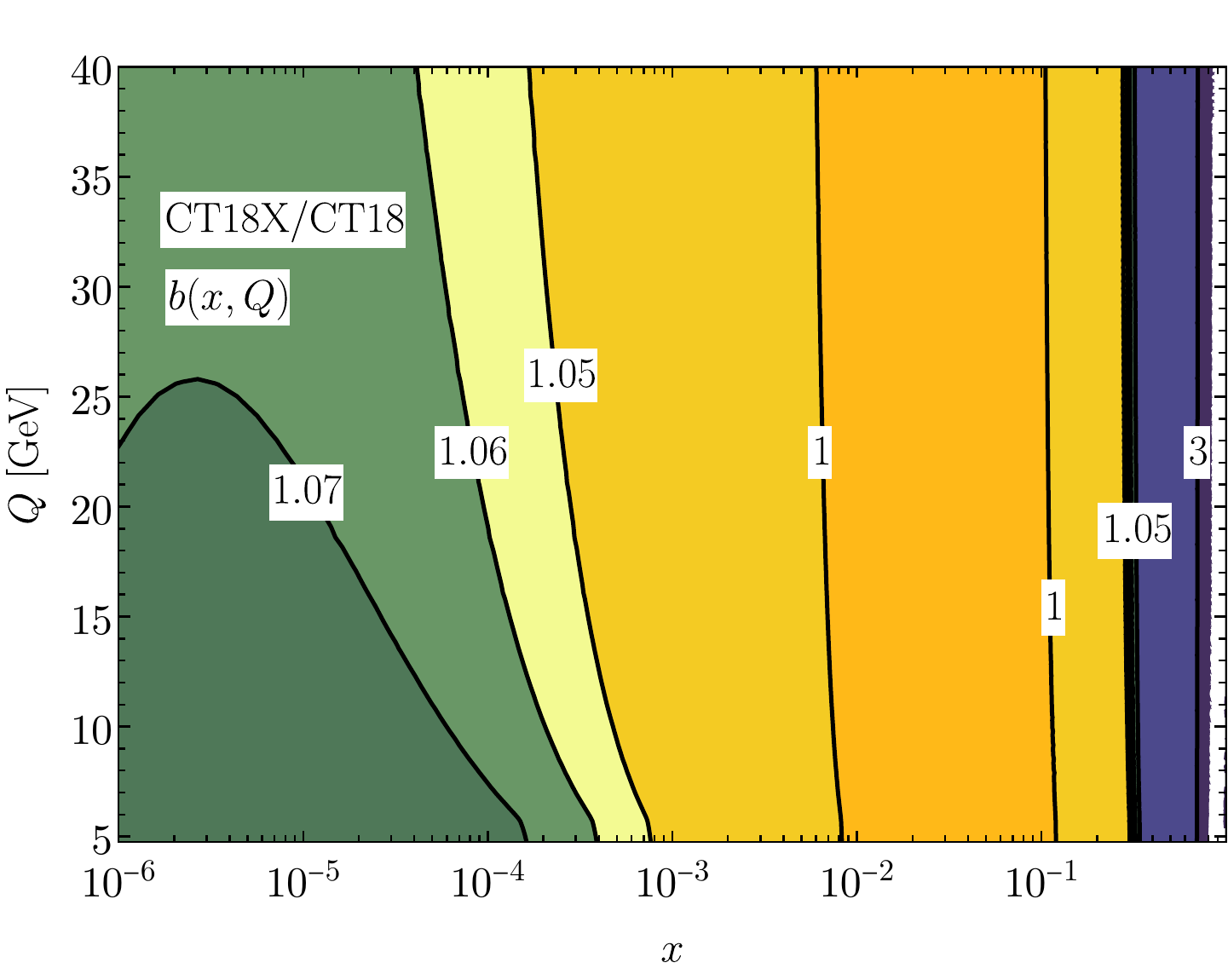}
\caption{CT18XNLO/CT18NLO PDF ratio for the gluon(left) and the $b$ quark(right).}
\label{fig:CT18X}
\end{figure}
For small $x$ below $10^{-4}$, higher-order QCD terms with $\ln(1/x)$ dependence grow quickly at low energy scales of order of few GeVs. This allows us to investigate a novel kinematic regime where both large- and small-$x$ QCD effects contribute to the $b$-hadroproduction rate. LHCb data can therefore be used to probe small-$x$ dynamics~\cite{Lipatov:1976zz,Fadin:1975cb,Kuraev:1976ge,Kuraev:1977fs,Balitsky:1978ic} and parton saturation~\cite{Mueller:2001fv}.

Recent global QCD analyses of PDFs have used different procedures to account for small-$x$ dynamics. For example, an improved DGLAP formalism has been used by the NNPDF Collaboration~\cite{Ball:2017otu,Ball:2021leu}, while the CT group introduced a variant of CT18 global PDF fit, named CT18X, that is generated by including an $x$-dependent scale choice for low-$x$ DIS data. This scale choice mimics the main impact of low-$x$ resummation and is motivated by partonic saturation models~\cite{Golec-Biernat:1998zce}. The two insets of Fig.~\ref{fig:CT18X}, show the differences between CT18 and CT18X PDFs expressed in terms of CT18X/CT18 ratios. In most of the $(x,Q)$ plane, the CT18X fit is very close to that of CT18, except for the small-$x$ (and small-$Q$) region where the gluon and $b$-quark PDFs are enhanced as a result of saturation. As seen Fig.~\ref{fig:PDF}, the $b$-quark density is highly correlated with the gluon. The main differences between CT18X and CT18 PDFs are in their corresponding uncertainties.
\begin{figure}
    \centering
    \includegraphics[width=0.79\textwidth]{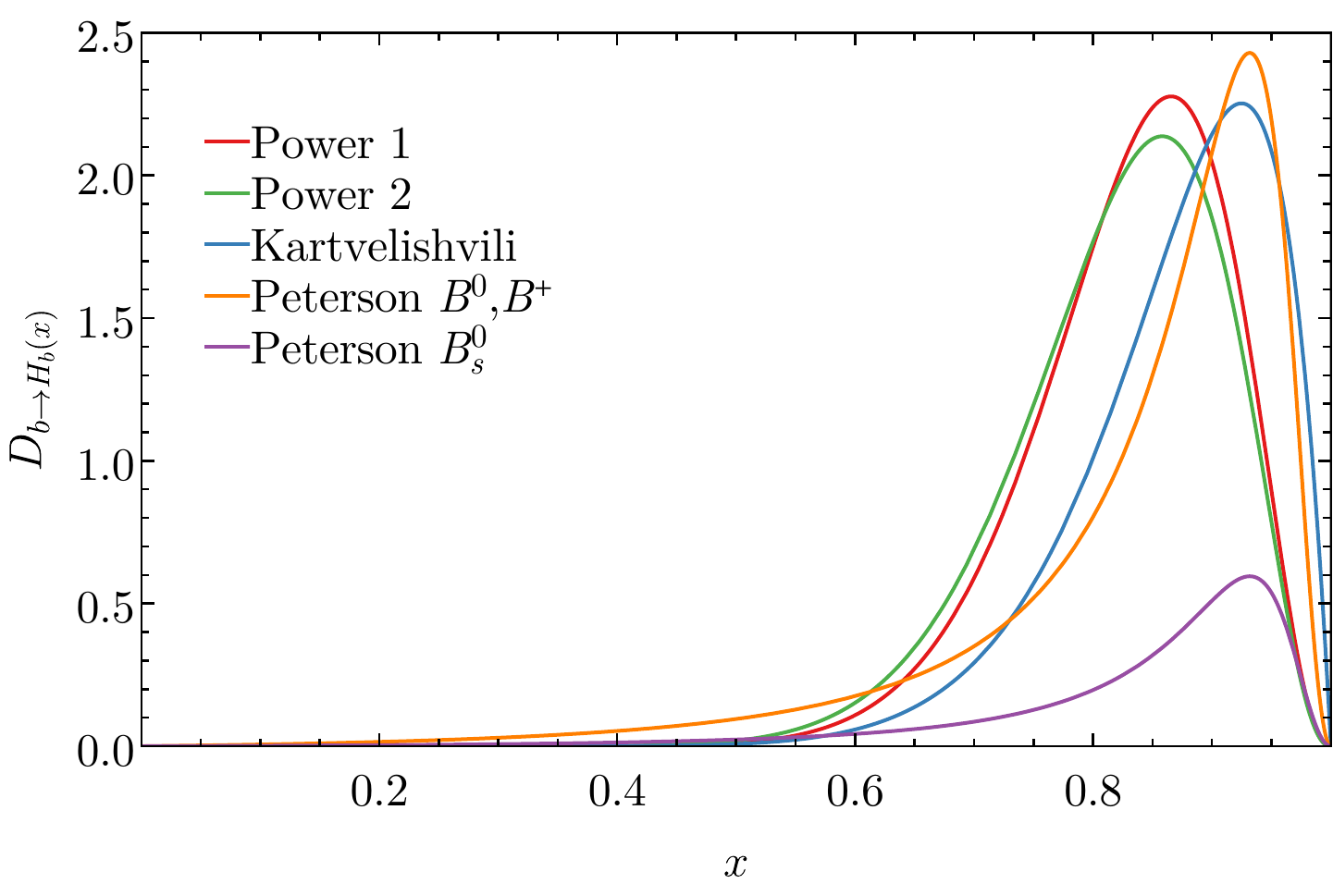}
    \caption{Fragmentation functions for $b\to H_b$, modelled according to the power ansatz described in~\cite{Salajegheh:2019ach}, and to the Kartvelishvili~\cite{Kartvelishvili:1977pi} and Peterson~\cite{Peterson:1982ak} parameterizations. The branch fraction is normalized to $\mathcal{B}(b\to B^0/B^+)=0.408$~\cite{ParticleDataGroup:2020ssz}.}
    \label{fig:FF}
\end{figure}

The LHCb cross section measurements for $b$-production~\cite{LHCb:2017vec} are presented at particle level and can therefore be used to probe $b$ fragmentation. In Fig.~\ref{fig:FF}, we illustrate various parametrizations for the $b$-quark FF $D_{b\to H_b}$. Labels ``Power 1 , 2'' refer to the power ansatz described in ref.~\cite{Salajegheh:2019ach}, while the Kartvelishvili and Peterson models are described in ref.~\cite{Kartvelishvili:1977pi} and~\cite{Peterson:1982ak} respectively. The branching fraction is normalized to
\begin{equation}
\mathcal{B}(b\to B^0/B^+)=\int\dd z D_{b\to B^0/B^+}(x)=0.408
\end{equation}
which is obtained from measurements of $Z\to b\bar{b}$ decays~\cite{HFLAV:2016hnz}.
The fraction measured at the Run II of Tevatron is slightly smaller ($\mathcal{B}(b\to B^0/B^+) = 0.340$~\cite{HFLAV:2016hnz}, but it has little if no impact on our results. A conservative estimate of the uncertainty associated to the FFs in this work is obtained by considering relative differences between the parametrizations mentioned above, where we have also included the Peterson model for the $b\to B_s^0$ fragmentation (see Fig.~\ref{fig:FF}), whose the corresponding branching fraction is normalized to $\mathcal{B}(b\to B_s^0)=0.100$~\cite{ParticleDataGroup:2020ssz}. A more rigorous estimate of FF uncertainties deserves a dedicated study which will be addressed in a future work.
\begin{figure}
    \centering
    \includegraphics[width=0.49\textwidth]{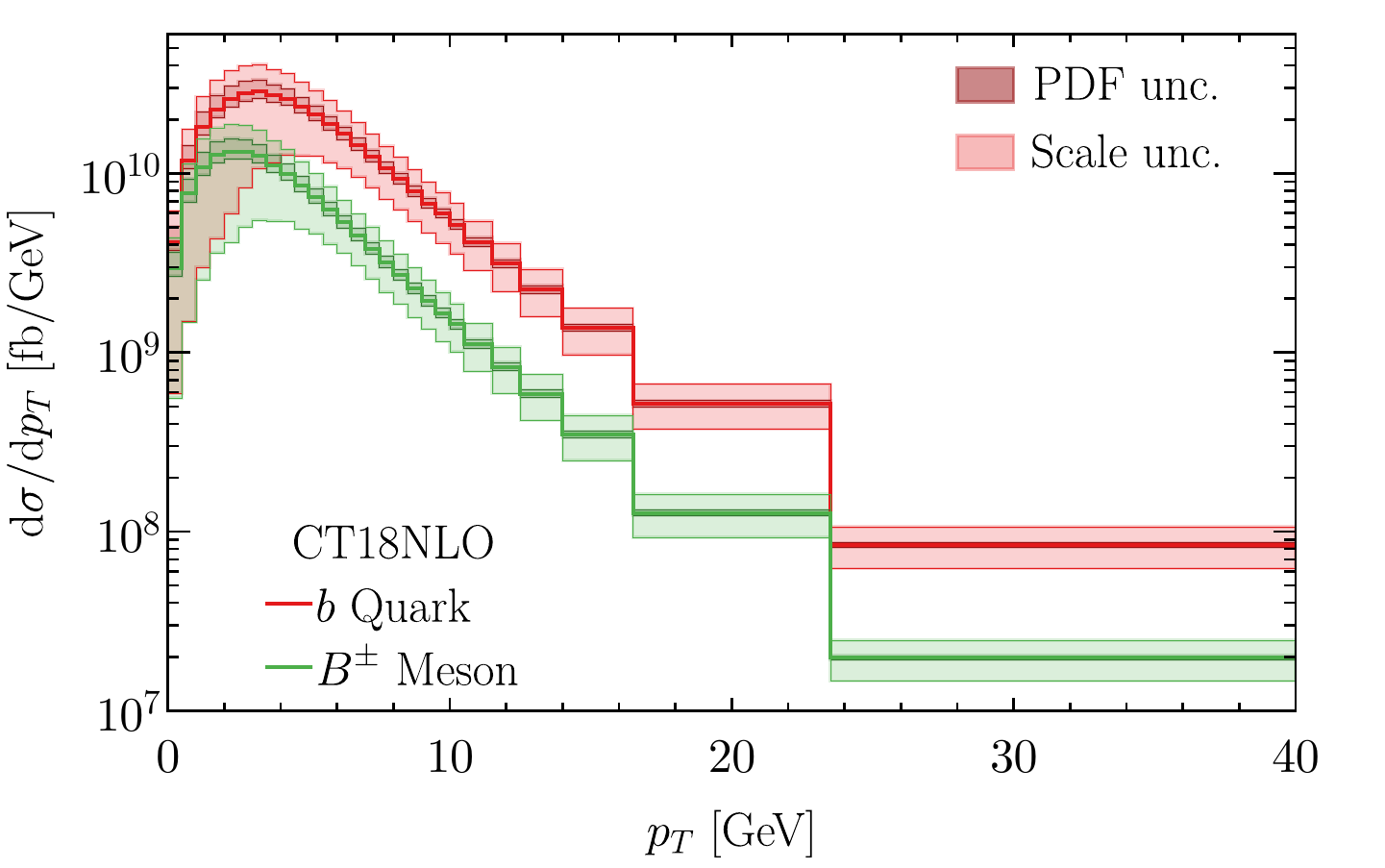}
    \includegraphics[width=0.49\textwidth]{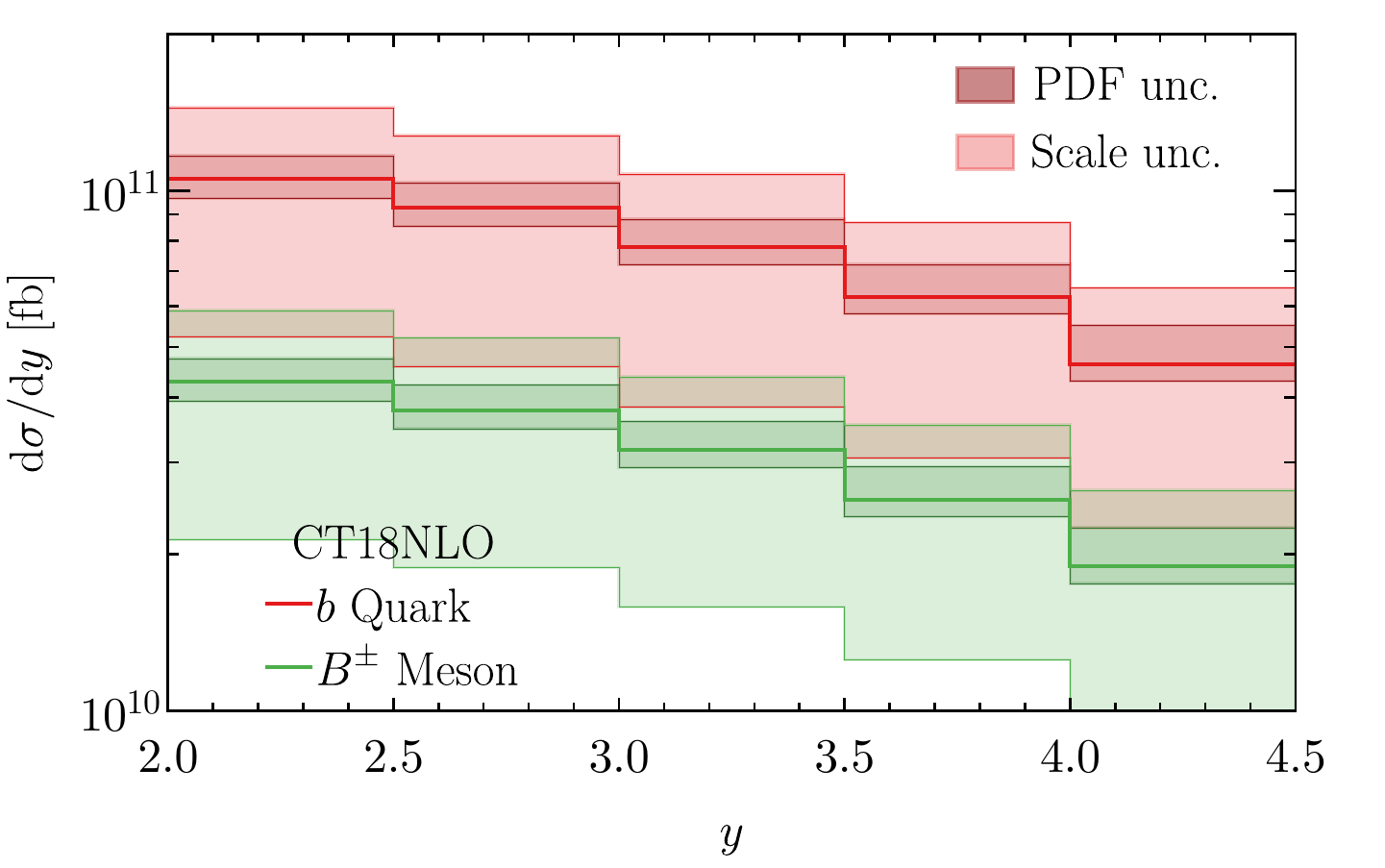}
    \caption{Data vs Theory comparison for the $p_T$ and $y$ distributions at parton and particle level for $b$-flavor hadroproduction at LHCb 13 TeV.}
    \label{fig:q2h}
\end{figure}

\section{Results for $B^\pm$ production at LHCb 13 TeV}

In this section we illustrate the phenomenological results of our analysis at NLO in QCD for $B^\pm$ meson production at the LHC 13 TeV. Cross sections are calculated by using the S-ACOT-MPS GMVFN scheme~\cite{Xie:2019eoe} and are compared to experimental measurements from the LHCb Collaboration published in ref.~\cite{LHCb:2017vec}.
In Fig.~\ref{fig:q2h} we illustrate a comparison between the theory predictions for the transverse momentum $p_T$ and rapidity $y$ distributions obtained at parton level and particle level. CT18NLO PDFs are used as input. Particle-level results are plotted in green while the parton-level ones are in red. The CT18 induced PDF uncertainties at the 90\% C.L. are represented by darker bands, while lighter bands represent scale uncertainty.
The central values for the renormalization ($\mu_R$) and factorization ($\mu_F$) scales are kept equal $\mu_R=\mu_F=\sqrt{m_b^2+p_T^2}$, and the scale uncertainty is obtained from the envelope of a 7-point variation by a factor of 2 of the scale ratio $\mu_R/\mu_F$. The induced PDF uncertainty on the cross section is approximately 10\%, as reflected by the gluon and $b$-quark PDF uncertainties in Fig.~\ref{fig:PDF}. Scale uncertainties are large and result from large variations in the running of the strong coupling $\alpha_s(Q)$, and from large variations of the gluon and $b$-quark PDFs in the corresponding $Q$ range.
In Fig.~\ref{fig:q2h}, the particle-level $B^\pm$ meson production $p_T$ spectrum is peaked at lower $p_T$ as compared to the parton-level distribution. This results from collinear fragmentation as the $B$-meson momentum is a fraction of the  $b$-quark momentum ($\vec{p}_{B}=z\vec{p}_b$). The rapidity distribution for the $B^\pm$ meson production appears to be approximately re-scaled by an overall factor $\mathcal{B}(b\to B^\pm)$ as compared to the $b$-quark production one.

\begin{figure}
    \centering
    \includegraphics[width=0.49\textwidth]{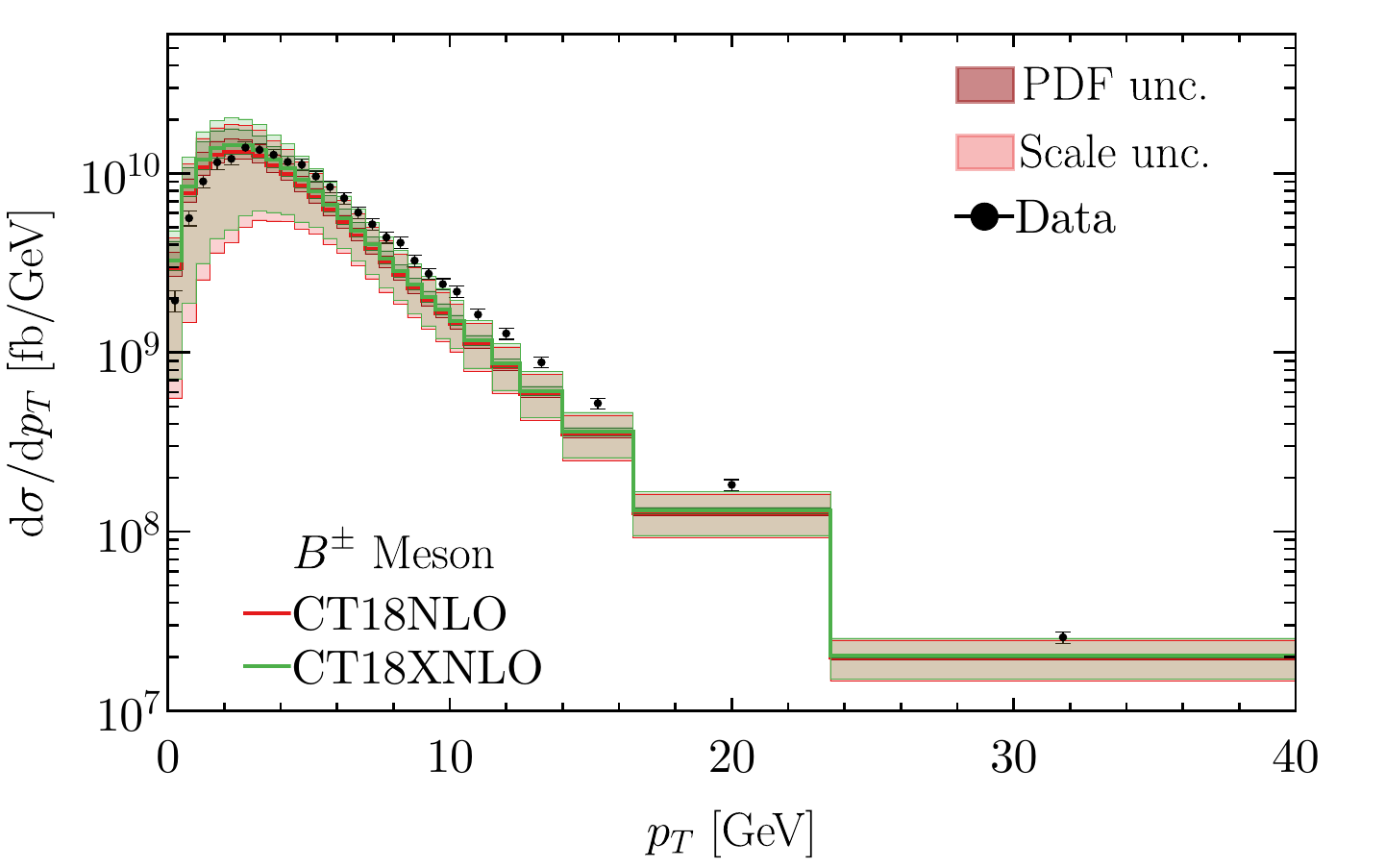}
    \includegraphics[width=0.49\textwidth]{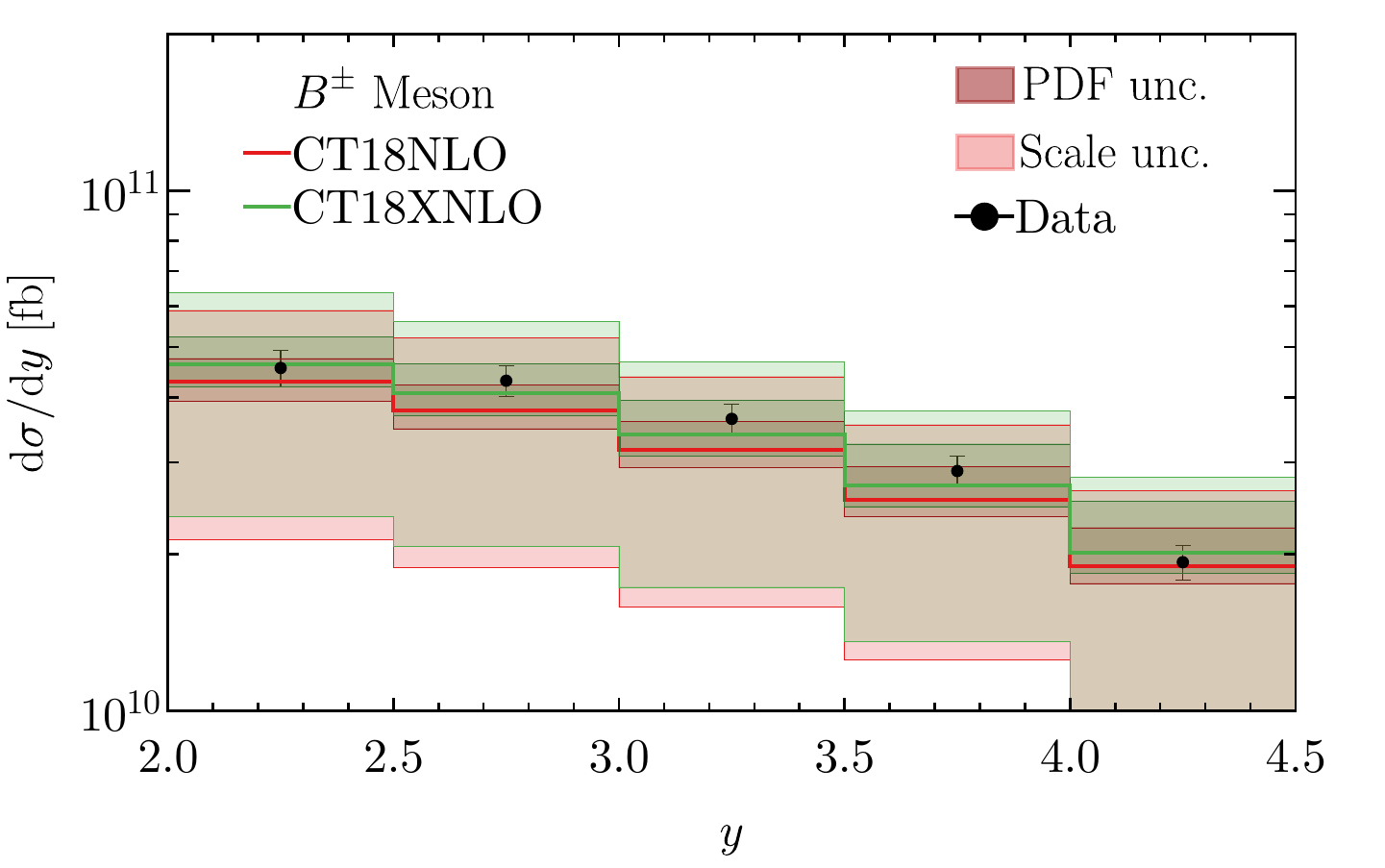}
    \caption{NLO theory predictions for the $p_T$ and $y$ distributions obtained with CT18NLO and CT18XNLO PDFs compared to the LHCb data for $B^\pm$ production at 13 TeV~\cite{LHCb:2017vec}.}
    \label{fig:hadron}
\end{figure}

In Fig.~\ref{fig:hadron}, we illustrate a comparison between our theory predictions for the $p_T$ and rapidity distributions for $B^\pm$ meson production with CT18 and CT18X PDFs, and the corresponding data from the LHCb experiment~\cite{LHCb:2017vec}.
The agreement between theory and data is overall good within the quoted uncertainties.
In particular, the predictions for the $p_T$ spectrum are in good agreement with data at low $p_T$. However, at $p_T > 10$ GeV data lie on the upper edge of the theoretical error bands. This suggests that higher-order corrections at next-to-next-to-leading order (NNLO) are necessary to improve the agreement between data and theory. The predictions for the rapidity distribution agree well with the data, although theory uncertainties are large. The rapidity central value obtained with the CT18X PDFs is in slightly better agreement with respect to that obtained with CT18. This is due to the enhanced CT18X gluon and $b$-quark PDFs as compared to CT18 (see Fig.~\ref{fig:CT18X}), which reflect small-$x$ dynamics effects captured by the CT18X PDFs.

\section{Conclusions and prospects}

In this study, we applied the S-ACOT-MPS GMVFN scheme to the production of $B^\pm$ mesons in proton-proton collisions. We compared theory predictions at NLO in QCD for the $p_T$ and rapidity spectra to precision measurements from the LHCb experiment~\cite{LHCb:2017vec} at a center of mass energy of 13 TeV. Overall, we find good agreement between data and theory. The LHCb experiment has the ability of measuring properties of $B$ mesons at large rapidity, and cross section measurements of $B^\pm$ meson production can be used to probe the $b$-quark and gluon PDFs at both small- and large-$x$ if consistently included in global QCD analyses of PDFs.
Theoretical uncertainties at NLO are large ($\mathcal{O}(50\%)$) and mainly ascribed to scale variation. This can be improved by including higher-order corrections which imply an extension of the S-ACOT-MPS scheme to NNLO. Observables with reduced theory uncertainties can in principle be obtained by considering cross section ratios. In this case, uncertainties cancel to a large extent between numerator and denominator due to $\alpha_s$ and PDFs correlations~\cite{Xie:2019eoe,Mangano:2012mh}. The S-ACOT-MPS scheme can easily be extended to other processes in proton-proton collisions that can probe heavy-flavor PDFs, e.g., $Z$ boson production in association with a $c$- or $b$-quark jet~\cite{Figueroa:2018chn}.
The inclusion of these processes in global QCD analyses will represent an important improvement for future PDF determinations that aim at reducing uncertainties of heavy-flavor PDFs, provided that a consistent general mass treatment for the case of proton-proton reactions is utilized.
In addition, a recent study based on cross section measurements at forward rapidity for $Z+c$ production at LHCb~\cite{LHCb:2021stx} has suggested a valence-like intrinsic-charm component in the proton wave function. This needs to be further explored in new global PDF analyses using precision measurements at the LHC that are sensitive to heavy-flavor PDFs, and a consistent general mass treatment to correctly account for mass effects.

{\bf Acknowledgments.}
We would like to thank D. Wackeroth and L. Reina for helpful discussions and communications.
The work of K.X. at is supported in part by the Department of Energy under Grant No. DE-SC0007914, the National Science Foundation under Grant No. PHY-1820760, and in part by PITT PACC. The work of M.G. is supported by the National Science Foundation under Grant No. PHY-2112025. The work of P.N. is partially supported by the U.S. Department of Energy under Grant No. DE-SC0010129.


\providecommand{\href}[2]{#2}\begingroup\raggedright\endgroup


\end{document}